\documentclass[letterpaper,twocolumn,10pt]{article}
\usepackage{usenix-2020-09}

\usepackage{amsmath,amsfonts}
\usepackage{algorithmic}
\usepackage{caption}
\usepackage{subcaption}
\usepackage{graphicx}
\usepackage{float}
\usepackage{textcomp}
\usepackage{xcolor}
\usepackage[dont-mess-around]{fnpct}
\usepackage{listings}
\usepackage{array}
\usepackage{booktabs}
\usepackage{appendix}
\usepackage{makecell}
\usepackage[switch]{lineno} 
\usepackage{comment}
\usepackage{multirow}
\usepackage{afterpage}
\usepackage{caption}
\usepackage{subcaption}
\usepackage{authblk}
\usepackage{xurl}
\usepackage[tight-spacing=true]{siunitx}
\newcolumntype{P}[1]{>{\centering\arraybackslash}p{#1}}
\newcolumntype{C}[1]{>{\centering\arraybackslash}m{#1}}

\title{\Large \bf That Escalated Quickly: An ML Framework for Alert Prioritization}

\author{Ben Gelman$^{*}$}
\author{Salma Taoufiq$^{*}$}
\author{Tamás Vörös}
\author{Konstantin Berlin}

\affil{Sophos Inc. \authorcr
  \{\tt ben.gelman, salma.taoufiq, tamas.voros, konstantin.berlin\}@sophos.com}

\begin{document}

\maketitle
\def\thefootnote{*}\footnotetext{These authors contributed equally to this work.}\def\thefootnote{\arabic{footnote}}
\begin{abstract}

In place of in-house solutions, organizations are increasingly moving towards managed services for cyber defense. Security Operations Centers are specialized cybersecurity units responsible for the defense of an organization, but the large-scale centralization of threat detection is causing SOCs to endure an overwhelming amount of false positive alerts -- a phenomenon known as alert fatigue. Large collections of imprecise sensors, an inability to adapt to known false positives, evolution of the threat landscape, and inefficient use of analyst time all contribute to the alert fatigue problem. To combat these issues, we present That Escalated Quickly (TEQ), a machine learning framework that reduces alert fatigue with minimal changes to SOC workflows by predicting alert-level and incident-level actionability. On real-world data, the system is able to reduce the time it takes to respond to actionable incidents by $22.9\%$, suppress $54\%$ of false positives with a $95.1\%$ detection rate, and reduce the number of alerts an analyst needs to investigate within singular incidents by $14\%$. 
\end{abstract}

\section{Introduction}
Starting at \$3 trillion in 2015, the cost of cybercrime has risen $15\%$ every year due to the increasing frequency, scale, and complexity of cyberattacks~\cite{cybercrime-cost2023}. A Security Operations Center (SOC), which is a specialized unit that monitors an organization's security posture and responds to security threats, is crucial to an organization's well-being. In the race to contend with the evolving landscape of cyberattacks, however, organizations are foregoing in-house security and subscribing to external services known as Managed Detection and Response (MDR) to protect themselves. 

In MDR, an external SOC provides comprehensive security for all of its customers, but the aggregation of so many diverse systems and use cases complicates analysis. To deal with the growing complexity of attacks, SOCs have employed increasingly numerous and complicated sensors such as network intrusion detection systems, antivirus, firewalls, log parsing software, and other similar solutions. These sensors generate a large number of alerts, overwhelming security analysts with redundant or false positive alerts -- a phenomenon known as alert fatigue. Analysts are obligated to inspect most of these alerts because ignoring them can have catastrophic outcomes. Such pressure can lead to organizational retention issues and increased operational costs~\cite{soc-study-2020}. Security information and event management (SIEM) platforms help streamline the process by partially normalizing data, but they still discard potentially useful data and ultimately fail to scale. 

\begin{figure*}[htbp]
\centering
   \includegraphics[width=1\linewidth]{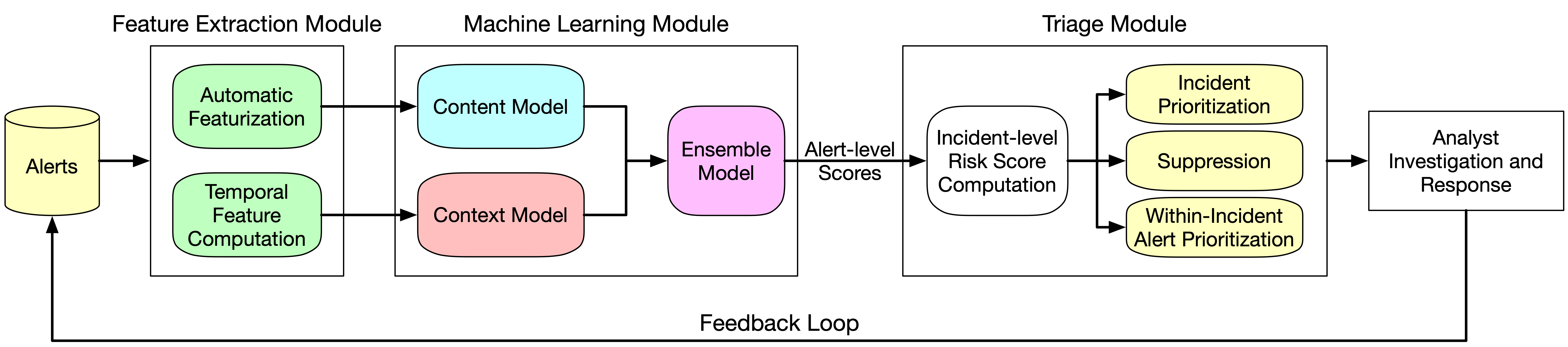}
   \caption{The TEQ system framework. Alerts pass through a feature extraction module comprising two components: an automatic featurization process that validates and includes data from semi-structured formats and a temporal computation process that examines historical context amongst customers and their endpoints. These feature vectors are passed to the machine learning module, which trains two separate models that are ensembled to create a final risk score. The triage module uses the risk score to provide actionable information to analysts in the form of incident prioritization, false positive suppression, and within-incident alert prioritization. The feedback loop automatically integrates analyst feedback into the system with minimal changes to manual workflows.}
   \label{fig:teq-overview} 
\end{figure*}

External SOCs are experiencing tremendous growth, with forecasts predicting a change in global market size from \$6 billion in 2022 to \$10 billion by 2027~\cite{socmarketshare}. This means that maximizing the efficiency of SOCs is a critical problem for the future of cybersecurity. Previous approaches have used numerous strategies, ranging from machine learning classifiers~\cite{alertrank-2021} and unsupervised learning~\cite{stream-clustering-2021} to effective GUIs~\cite{awalin-2018} and theoretical solutions~\cite{game-theory-1-2017, game-theory-2-2017}. Common oversights in the application of these systems are the potential for a wider diversity in data collection, the ability to leverage as much data as possible from multiple organizations, the evolution of data over time, and the continuing role of human analysts. We focus our attention on four main contributors to alert fatigue: large collections of imprecise sensors, an inability to adapt to known false positives, evolution of the threat landscape, and inefficient use of human analyst time. In this work, we present That Escalated Quickly (TEQ), a machine learning framework for alleviating alert fatigue with minimal changes to existing workflows in the context of a real-world SOC. Figure~\ref{fig:teq-overview} displays the four modules of TEQ, which address the above issues plaguing SOC efficiency. We explain the problems in more detail below and describe how TEQ contributes a solution.

First, sensors are inconsistent in the data they collect and the way that data is formatted. Employing multiple such sensors potentially compounds the alert problem~\cite{alert-fatigue-ai-2021}. It also results in missing data and high maintenance or update costs. TEQ's feature extraction module automatically parses semi-structured data from arbitrary collections of sensors, handles missing values, and validates the feasibility of inputs to machine learning models beyond the standard processing of SIEM platforms. 

Next, false positives are frequently duplicates or near-duplicates of alerts that analysts have already resolved. Blocking all possible variations of a false positive is infeasible, and manual efforts to blocklist these alerts are time-consuming and can usually only handle a few of the most common circumstances. TEQ's machine learning (ML) module learns to identify false positives using a data-driven approach, and it is robust to many near duplicates because the models do not use hard-coded rules to make their predictions.

Subsequently, new threats are constantly emerging, and encoding domain knowledge into existing sensors is a tremendous effort. The tuning required to identify new threats in itself is not a trivial task, as domain experts need time to understand and delineate those threats. TEQ's feedback loop is a non-intrusive method for integrating new knowledge into the system. When analysts manually resolve an incident, which is an alert grouping comprising all of the alerts and information related to an instance of malicious behavior, the outcome is automatically propagated to the constituent data. The data is reprocessed through the feature extraction and machine learning modules; thus, even if the analysts do not write any additional domain knowledge rules in response to the new threat, the ML models learn to associate existing sensors with the new data. This works because sensors can contain information that detects multiple threats. This outcome-based feedback loop does not require any additional effort on the part of analysts beyond their standard incident resolution.

Finally, presenting knowledge to analysts efficiently is crucial for their productivity. When an analyst resolves an incident, the team may not have direct insight into what motivated the solution. This can result in analysts repeating work for false positives, similar incidents, or new attacks. TEQ's triage module uses three strategies to actualize the ML module's knowledge and improve analyst efficiency with minor changes to existing workflows. Incident prioritization reorders an analyst's incident queue, bubbling up incidents that are most likely to be malicious. Suppression is an extension of that reordering, providing a single, tuneable threshold to eliminate the most likely false positives at the bottom of the queue. Within-incident alert prioritization focuses analyst attention on the most important data within a single incident; thus, even if an incident requires manual resolution, analysts can spend less time determining the outcome. With these three methods, the only addition to the existing workflow is the score from the machine learning ensemble, which ultimately manifests as a reordering of elements the analysts are already familiar with. 

In summary, our contributions are as follows:
\begin{itemize}
  \item Demonstration of feasibility of a hands-off featurization system that handles semi-structured data from arbitrary sensors without hindering downstream tasks.
  \item An ensemble of models that operate on a greater breadth of alert and temporal features than have been explored in previous security datasets.
  \item A novel, in-depth evaluation of alert prioritization performance and feature importance over time, demonstrating the changes and continuities across an evolving threat landscape.
  \item The first system, to our knowledge, that utilizes both alert-level and incident-level scores, allowing for standard incident prioritization and the ability to identify key information within a single incident. 
  \item A simple triage system that shows $22.9\%$ reduced queue times for actionable incidents, a $54\%$ overall false positive suppression rate, and $14\%$ reduced incident resolution time on a deployment scenario simulated on real data from a beta version of an MDR product from a large security company. 
\end{itemize}

The rest of the paper is organized as follows. Section~\ref{sec:related} presents a review of related work. Section~\ref{sec:methodology} details our methodology and design decisions based on the nuances of our real-world alert data. Section~\ref{sec:experimentsresults} then describes our experimental setup and demonstrates our results in realistic scenarios, along with analysis and discussion. Section~\ref{sec:conclusions} expresses final thoughts.

\section{Related Work}
\label{sec:related}
A wide variety of strategies have been applied to alleviate alert fatigue via triage and prioritization modeling. Supervised machine learning is of particular interest to our work; prior research has demonstrated the potential of these techniques in combating alert fatigue~\cite{dl-prioritization-2017, user-centric-soc-2017, fp-reduction-ids-2016, alertrank-2021, learning-to-rank-2016, alert-graph-2019, alert-fatigue-ai-2021}. For instance, the authors of~\cite{alertrank-2021} propose an alert ranking solution in the context of online service systems. They train an XGBoost ranking model on various alert features (textual, temporal, univariate, and multivariate anomaly-based) to identify severe alerts for human security operators. However, this existing research is limited to applying classification algorithms to fixed alert data. The incorporation of expert knowledge and feedback is generally briefly addressed. In our proposed work, we include a feedback loop to utilize security analysts’ expertise and create a holistic solution to the problem.  

The system proposed in~\cite{awalin-2018} incorporates a feedback loop. The authors present a three-part solution: a random forest classifier trained on features extracted from alerts' contents (e.g. command length, process tree, occurrence of certain characters); a prediction view presenting the model’s decision and an explanation; and a dashboard providing an analysis of the model’s performance. Their feedback loop allows analysts to provide a decision in the prediction view, which is then fed back to the model. TEQ also uses a feedback loop to integrate analysts' expertise, but, in contrast, we further refine our system’s adaptability to account for the constant changes in underlying sensors with an automatic featurization of alert contents, rather than a manual feature extraction procedure. Another notable difference is that our system employs an ensemble of machine learning models on different sets of signals, using not only alert contents but also their temporal firing patterns.

Other techniques used to model the problem include unsupervised learning with the application of clustering algorithms~\cite{stream-clustering-2021, fp-ids-clustering-2015}, isolation forests~\cite{isolation-forest-sae-2020}, or other unsupervised algorithms~\cite{nodoze-2019, reinforcement-learning-2020} and non-machine learning methods. For instance, some works have adopted game theoretic approaches: in~\cite{game-theory-1-2017, game-theory-2-2017}, the authors treat alert prioritization as a Stackelberg game with the underlying assumption that attackers have intimate knowledge of the defense/detection strategy. The authors formulate an NP-hard problem and demonstrate approximate solutions. Although these are interesting theoretical approaches, the approximations use a more rigid attacker dynamic and still have difficulty scaling to massive amounts of data.

As mentioned earlier, a crucial contribution of our work is the automation of the featurization process specifically on semi-structured alert data. A variety of existing work and libraries have tackled data parsing~\cite{unstructured-iot, semistructured-ml, autojsonparsing, dynamicparsing} and model optimization~\cite{automl-list}, but automated feature extraction has either been limited to specific domains, such as physiological signals~\cite{pysiology} or electrodermal activity~\cite{electrodermal}, or has some expectation of a predefined schema in the case of more general frameworks~\cite{tsfresh, featuretools, dfs, autofeat, autofeature-list, feature-general-constructor, cognito}. Our work accommodates the featurization of semi-structured data with missing data and a partially-defined schema. 

\section{Methodology}
\label{sec:methodology}

In this section, we explain the design decisions for each component of the TEQ system shown in Figure~\ref{fig:teq-overview}. At the highest level, the diagram reflects a standard process in machine learning: data goes through feature extraction to train machine learning models whose output scores are presented to humans; however, the complexities of the cybersecurity domain necessitate a nuanced approach, which we elaborate on throughout this section.

\subsection{Data}
\label{sub:data}

\begin{table*}[!t]
\centering
\begin{tabular}{|p{0.08\linewidth} | p{0.15\linewidth}  | p{0.21\linewidth} | p{0.38\linewidth} |} 
\hline
Field & Description & Alert 1 & Alert 2 \\
\hline
Alert \newline Signature & Signature that \newline triggered the alert &  POWERSHELL-80606b5bc125ce99b189731 & CLEAN-ATK-Mimikatz-AS \\
\hline
Command Line & Executed \newline command line & 
 \texttt{powershell Start-BitsTransfer -Priority foreground -Source <url> -Destination C:\textbackslash Windows\textbackslash Temp\textbackslash}
 
 \texttt{bitsadminflag.ps1}
&  N/A \\
\hline
Filepath & Path of the file that triggered the alert & N/A  & 
\texttt{C:\textbackslash Users\textbackslash<user>\textbackslash tools\textbackslash Covenant\textbackslash}

\texttt{Data\textbackslash EmbeddedResources\textbackslash}

\texttt{SharpSploit.Resources.powerkatzx86.dll}
\\ 
\hline
\end{tabular}
\caption{Three fields from alerts generated by different sensors. The first alert is triggered by a suspicious command via the \texttt{Start-BitsTransfer} cmdlet, which could be used to download, execute, and clean up malicious code. The second alert was triggered by a signature for Mimikatz, which is an exploit on Microsoft Windows that extracts passwords stored in memory.
}
\label{tab:alert-examples} 
\end{table*}

To be more specific, our approach works by predicting a score for an \emph{alert}. An alert is an abstraction for any piece of information that is indicative of malicious behavior on a device. These data are retrieved by \emph{sensors}, which are mechanisms that are checking for malicious activity, including antivirus software, firewalls, and hand-written rules that read logs or create regular expressions to analyze anything happening on a device. Therefore, alerts can contain information about any security related event, such as a file accesses, command line executions, network communications, etc. Experts have spent decades analyzing that data to identify patterns, determine signatures, and draft rules to automate detection. It is not the goal of our system to replace domain expertise, but rather to work in concert with experts to maximize the performance of the SOC as a whole. Thus, the first crucial characteristic of our approach is that we are operating on severe alerts rather than all events. Every alert in our data set is considered severe enough by the author of the sensor to require an investigation from human analysts.

The individual alerts in our system are represented as semi-structured collections of data from customer endpoints that are acquired via sensors and stored in JSON files. The data is partially normalized across alert types by a SIEM platform, such that common fields like \texttt{machine id}, \texttt{alert severity}, \texttt{alert type}, various timestamps, and some other basic information is located under the same JSON path. Outside of those fields, however, the schemas are vastly different between alerts. This strategy provides a highly flexible means to add, change, or remove sensors without significant time investments in modifying infrastructure, meaning that analysts may modify sensors quickly without concern for breaking downstream pipelines and storage systems. On the other hand, the resulting diversity in the alert schemas creates numerous issues for downstream, automated analysis: 1) there are too many fields to manually codify, and SIEM platforms only retrieve common data, leaving many fields that are unused or added/removed over time; 2) there are constantly missing values because an alert generated by one sensor naturally will not have the same fields as an alert generated by another sensor; and 3) the data can be arbitrarily nested within multiple levels, with no definition of that structure given before attempting to parse it. 

\begin{figure*}[!ht]
\begin{subfigure}{.5\textwidth}
  \centering
  % include first image
  \includegraphics[scale=.498]{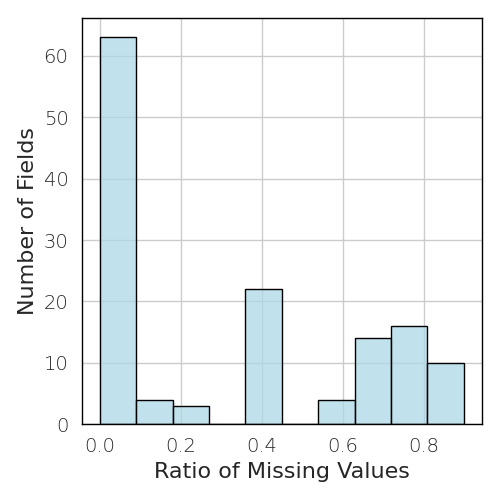}  
  \caption{Missing field ratios.}
  \label{fig:missing-features-histogram}
\end{subfigure}
\begin{subfigure}{.5\textwidth}
  \centering
  % include second image
  \includegraphics[scale=.5]{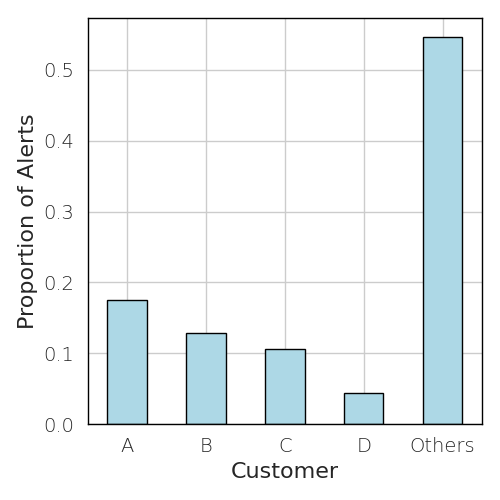}  
  \caption{Alert distribution across customers.}
  \label{fig:customer-rep}
\end{subfigure}
\hfill
\newline
\hfill
\begin{subfigure}{\textwidth}
  \centering
  % include third image
  \includegraphics[width=0.85\linewidth]{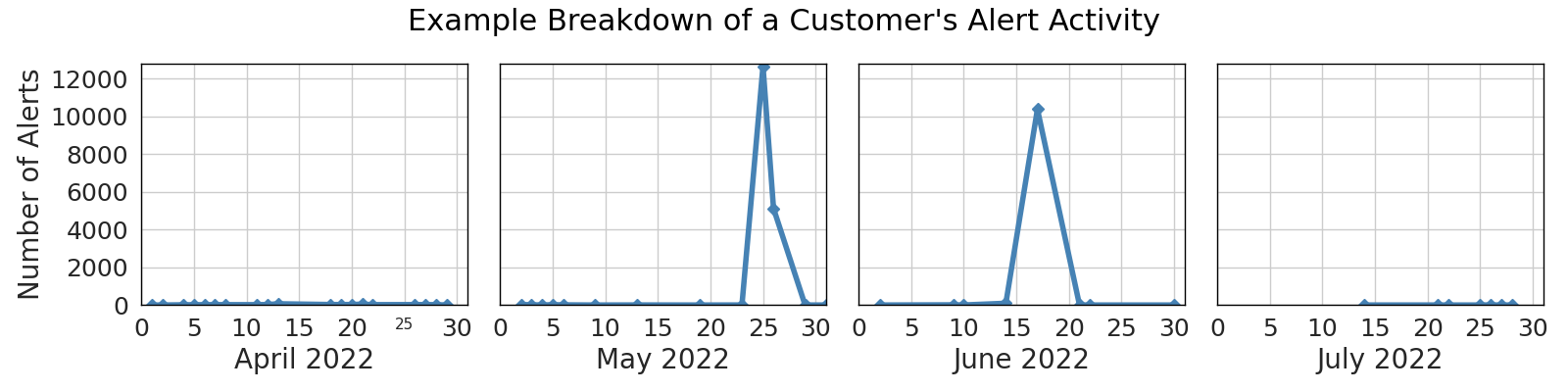} 
  \caption{Monthly breakdown of sensor activity on a customer estate showcasing three different types of alert activity: zero activity, a low, steady baseline, and significant bursts.}
  \label{fig:alerts-monthly-cust}
\end{subfigure}
\caption{Examples of alert data characteristics.}
\label{fig:data-characteristics}
\end{figure*}

Table~\ref{tab:alert-examples} displays a subset of two example alerts: one where a suspicious command line downloads a shell script from the internet and executes it in-memory, and one where a portable executable (PE) is identified as an exploit (Mimikatz). When tabulated, the alerts have missing data because the keys did not occur in their JSON files. Figure~\ref{fig:missing-features-histogram} further illustrates the magnitude of the missing data problem by emphasizing the quantity of fields that have a significant proportion of missing values. Due to the diversity of sensors, over $70$ fields are sparsely populated. We also observe that there are $63$ fields with a very low missing ratio. These fields are likely to contain more generic endpoint data or meta information such as customer and machine identifiers or timestamps. To alleviate these issues, we create an automatic featurization framework, which we explain further in Section~\ref{sub:autofeature}. 

The data from alerts, however, is not the only source of information by which we can solve the alert fatigue problem. It makes sense to gain an understanding of customers' behaviors, even if that data does not come from a specific sensor, because we are trying to maximize the efficiency of the SOC as a whole. As a result, we also capture temporal information about customers, machines, and sensors. We elaborate on the purpose and utility of this data in Section~\ref{sub:temporalfeatures}.

With the alert data structure defined, the subsequent notion of \emph{incidents} is critical to the functionality of the framework. An incident is a group of alerts that tend to relate to a singular security event. Because each alert can contain different data based on the sensor from which it originated, creating an incident is a convenient abstraction for a holistic analysis of a security event. For our SOC of interest, an incident is created using a straightforward strategy that groups all alerts that happen within 24 hours on the same machine. Prior work has used a variety of alternative grouping strategies, but we consider grouping optimization outside the scope of our current work. 

The incident level is where human intervention begins. Human analysts must investigate and respond to each incident, effectively addressing all alerts that have been grouped. It is at this incident level that analysts determine whether an incident was a false positive, thereby assigning a label. This incident labeling is the core principle of TEQ's feedback loop because a manageable amount of incidents can be manually resolved, propagating the label back to an unmanageable amount of alerts. 

\begin{table*}[!ht]
\centering
\begin{tabular}{|p{0.81\linewidth}|p{0.14\linewidth}|} 
\hline
Incident Description & Actionable Label\\
\hline
    The SOC team investigated a detection for ProxyShell exploitation and LemonDuck malware. We identified the anti-virus solution is continually detecting and cleaning webshells from the host. The host is still not patched to the latest Exchange version. \textbf{We recommend completing the recommendations in order to prevent the server from being exploited and to remove the LemonDuck persistency. Escalating to client due to to Lemonduck persistence.} &  True \\
\hline 
    The file \texttt{C:\textbackslash{Users}\textbackslash \textless{USER}\textgreater\textbackslash AppData\textbackslash Roaming\textbackslash7kA3320\textbackslash UxTheme.dll} \textbf{has been cleaned} by the anti-virus solution and \textbf{the file was not executed} on the host. We found no suspicious activity on the host. \textbf{No further actions are required}. & False \\
\hline
    The SOC team received a no VBS extension alert. We have observed this activity to be related to FPS PDF Driver. As a precaution we have checked running processes and persistence items such as scheduled tasks, start-up items and run keys with nothing of note found. \textbf{No action is required at this time.} & False\\ 
\hline
\end{tabular}

\caption{Incident labeling examples for the actionable label.
} % title of Table
\label{tab:incident-labeling} 
\end{table*}

The nature of the alert fatigue problem, however, requires a nuanced change to the data labels. TEQ is not intended to simply detect malicious activity. The sensors, such as antivirus software and firewalls, are already doing that to generate the alerts the analysts encounter. Instead, TEQ aims to answer the question \emph{``Does this alert require human intervention?''} We dub this interpretation the \emph{actionable} label, indicating whether an analyst has to take action to resolve the problem. This difference ensures that the prioritization system bubbles up alerts that cannot be resolved without analyst attention. The actionable labels are assigned based on the following criteria:

\begin{itemize}
  \item Incidents that require any kind of manual remediation are labeled as positives.
  \item Incidents generated by true positive alerts, i.e. malicious activity occurs, but is successfully contained by automated defense infrastructure, are labeled as negatives.
  \item Incidents generated by false alerts, i.e. no  malicious activity occurs, are labeled as negatives.
\end{itemize}

The second bullet may seem counter-intuitive, but as stated before, it is not TEQ's goal to simply detect malicious behavior. Requiring an investigation from human analysts every time an antivirus cleans an actual threat, for example, is still a waste of time. This is clearer in Table~\ref{tab:incident-labeling}, which provides examples of real incidents for each of the labeling categories. 

The following section describes how we perform feature extraction in light of the issues and complexities of the data we have identified here.  

\subsection{Feature Extraction Module} 
\label{sub:featureextractionmodule}
Based on the qualities of the data discussed in Section~\ref{sub:data}, we extract two distinct sets of features. The first set deals directly with the contents of alerts, while the other gathers context around behavioral trends. 

\subsubsection{Automatic Featurization}
\label{sub:autofeature}

To reiterate, even with the partial normalization of a SIEM platform, there are frequent missing values, unknown data structures in the data source, and too many fields to manually codify. To automatically adjust for these challenges, we use an automatic featurization step. Our framework analyzes the types, quantities, and distributions of the raw alert contents in order to construct and refine feature vectors that are provided to the model as input. This allows the system's training and deployment pipeline to adapt to new alert types and schema changes without human intervention.

The automatic featurization framework uses the following steps to convert semi-structured alert JSONs to validated feature vectors. First, the framework accepts arbitrarily structured JSONs and follows a flattening procedure to identify all of the keys. Nested keys are prefixed by the key in which they are contained. Arrays are treated as strings. All unique keys are tracked throughout the parsing procedure, and a NaN is assigned as the value for any key that does not exist within a given JSON file. This results in a sparse table where each row is an alert. The columns are defined by the unique keys, populated by the available values that are in the alert JSONs and NaNs for the missing values. 

Next, we perform validation on the data to ensure that it is a viable input to a machine learning model. We first automatically remove all columns whose names contain identifier language such as \texttt{id}, \texttt{time}, or \texttt{epoch}. Due to the partial normalization of the SIEM platform, this ensures that a variety of unique identifiers and timestamps are removed before training, which is useful because these features do not generalize to test data. Although it is possible to drop useful, but poorly named, columns, the amount of data that remains is still significantly larger than the amount of data that results from the partial normalization of the SIEM platform. We then try to parse columns into floats or ints. The NaNs that we use to represent missing values during parsing are replaced with $-1$ to avoid mixing data types and colliding with existing values, though a collision may still occur given the unknown nature of any particular column. The rest of the columns are treated as strings and undergo more processing.

All of the remaining string fields have their NaNs replaced with the string ``missing\_val'' to once again unify the data type within the column. Because strings can have an immense long tail of possible values, we need to ensure that the data will be useful when training the machine learning models. To do this, we first compute the count of unique values in each column. We then use a tuneable threshold to replace any string that falls below the threshold with the string ``rare\_val.'' Although this may remove some information from the column, it also prevents noise from flooding the model during training. Additionally, the ``rare\_val'' string often becomes an indicator for an anomalous reading. Empirically, we find that a threshold of 50 performs well, but the threshold can be easily tuned along with other model hyperparameters. Finally, we perform one last uniqueness check that removes any string columns that only have one or two unique values. This removes columns that only contain the ``missing\_val'' and ``rare\_val'' strings. The remaining strings are one-hot encoded; the substitution of the ``rare\_val'' for all strings under a certain threshold prevents the number of encodings from growing out of control. For our data, the entire procedure results in 101 alert content features. 

\subsubsection{Context Feature Computation}
\label{sub:temporalfeatures}

To elaborate on the temporal data discussed in Section~\ref{sub:data}, the \emph{context} around an alert, such as the knowledge that the customer where the alert originated has received a hundred alerts in the last week, can influence the manner in which analysts should interpret the alert. Figure~\ref{fig:customer-rep} shows that alerts are not uniformly distributed across customers. Prevalent customers can drown out signals from smaller customers when considering only the content of the alert data. Furthermore, Figure~\ref{fig:alerts-monthly-cust} demonstrates that breaking up the alert activity over time for individual customers uncovers different behavior patterns.

Context feature computation aims to track a meaningful variety of signals across the entirety of the SOC user base, which comprises a diverse set of organizations and endpoints. We design features to capture real-time trends, such as customer estate size, customer and endpoint vulnerability, alert activity patterns, and sensor activity. Several temporal features and summary statistics over numerous granularity time windows are used to encapsulate these behavioral signals. This data is capable of capturing information such as sudden anomalous peaks in alert volume that may indicate sensor noise or misconfigurations.

To be more specific, features for the context-based classification are computed across various predicates over nine different time windows ranging from one minute up to one week, culminating in a total of 90 numerical features. The time windows are of lengths: $60$; $120$; $300$; $600$; $3,600$; $7,200$; $43,200$; and $60,4800$ seconds. Experiments have shown that shorter time windows do not add significant information. The largest window is one week in order for the look-back feature computations to be tractable. Additionally, this window size allows for approximating the overall customer estate size and sensor firing activity. The following list shows the predicates:
\begin{itemize}
    \item Count of alerts on customer's estate.
    \item Count of alerts on endpoint.
    \item Count of alerts fired by a sensor.
    \item Average alert severity scores for alerts fired for an endpoint.
    \item Average alert severity scores for alerts fired by a given sensor.
    \item Count of sensors that have fired on a customer’s network.
    \item Count of customer endpoints that triggered alerts.
    \item Count of endpoints for which a sensor has fired.
    \item Count of customers on whose estates a sensor has fired.
    \item Count of alerts triggered on a customer's network by a given sensor.
\end{itemize}

\subsection{Machine Learning Module}
\label{sub:mlmodule}

TEQ's ability to combat a rapidly changing threat landscape depends on an efficient training/test cycle and a wide net of features and modeling strategies. It is not necessary for a single model or feature to invariably solve the prioritization problem. As attackers change strategies, data evolves, and the model/critical features best suited for the moment may change from week to week. By having the freedom to choose from a wide breadth of features and a diverse set of models, TEQ can make the adaptations required to perform well even in unexpected or specific circumstances.  

With that in mind, we consider four algorithms for our machine learning models: logistic regression (LR) for a baseline linear method, random forest (RF)~\cite{randomforest} and XGBoost (XGB)~\cite{xgboost2016} for tree-based methods, and a feed-forward neural network (NN) for a more general non-linear model. These algorithms are fast to train even on low-cost compute resources and traverse the optimization space in sufficiently different ways. As a result of the two features sets described in Section~\ref{sub:featureextractionmodule}, we likewise have two models that use these four algorithms. 

The content model acts directly on the alert features from the automatic featurization framework and the context model acts on the features from the temporal feature computation. In order to combine these two models, we use ensembles with four different statistical aggregations:
\begin{itemize}
    \item The average of the predictions.
    \item The maximum of the predictions.
    \item The weighted average with 70\% weight for the content model and 30\% for the context model.
    \item The weighted average with 30\% weight for the content model and 70\% for the context model.
\end{itemize}
We choose to use these statistical aggregations because they can adjust for the success or failure of a particular model, and they are efficient because they do not require retraining or calibration data. The full training procedure thus creates four content models, four context models, and 64 ensembles. Before discussing model selection, we will echo the distinction between alerts and incidents and explain training and test splits. 

As we discussed in Section~\ref{sub:data}, incident labels are propagated back to label their constituent alerts. This feedback strategy allows the model to train at the alert level, which is important because we do not have to aggregate alert features in order to train on an entire incident at once. However, we still desire a model that performs the best at the incident level. The models, having been trained on alert-level data, output alert-level scores from $0$ to $1$, indicating the actionability of an alert. Subsequently, we convert alert scores to incident scores by taking the maximum model prediction of all alerts within the incident. We refer to the resulting incident-level scores as TEQ incident scores. This conversion allows us to then validate the models against the incident-level labels. 

Beyond the conversions between alerts and incidents, an absolutely vital component to successfully training and evaluating the machine learning models is the use of a time split in the data. Whenever the passage of time can cause fundamental changes in the domain, such as the creation of a new cyber attack, then a random split in the data can have catastrophic effects on the model because it can overfit on information from the future that it would never realistically see at the time of training. Thus, we use six months of data, where the first 5 months are for training and the last month is held out for testing. 

Combining the incident-level evaluations and this month of held-out test data, we are able to perform standard model selection by choosing the model with the best test set performance at the incident level. Although it may seem like the traditional validation split for hyperparameter tuning is missing, the justification lies in the temporal nature of the data. Because the models are indeed sensitive to how recently they have been trained, which we demonstrate in Section~\ref{subsec:decay}, further splitting the time-ordered training data to use as a validation/hyperparameter tuning set causes a significant decrease in performance on the test set. Although we do try a variety of hyperparameters with that extra validation data split from the training data, we find that conservatively selecting the default hyperparameters from the Scikit-Learn~\cite{sklearn} library yields the best results over time, though the size of the data could potentially change this in the future.

\subsection{Temporal Analysis and Feature Interpretability}
\label{subsec:temporal-analysis}

With a trained ensemble from the machine learning module, we are able to perform model evaluations. For the purposes of this work, we take an extra step to evaluate the performance of our system over a significant period of time and to determine how the features important to TEQ's performance may change over that time. In order to do so, we include two evaluations that we call the \emph{decay} experiments. Typically, we train and test the model on 6 months of data as described in Section~\ref{sub:mlmodule}; however, the temporal nature of the data raises a question about what happens to the performance of a model with and without frequent retraining. Our two additional experiments attempt to answer that question.

For both decay experiments, we retrieve an extra two months of data prior to the start of our original set, resulting in eight months total. In the first experimental scenario, we train a model on a \emph{fixed} training set corresponding to the first five months, test on the subsequent sixth month, select the best performing content, context, and ensemble models, and then evaluate those on the remaining two months. This procedure provides insight into how the performance decays over time.

In the second experiment, the system is retrained by \emph{sliding} the training window in order to learn whether updated data prevents performance degradation. The sliding window cuts a six-month dataset from the eight months available, runs training on the first five and tests on the sixth, then slides by one month for retraining and re-evaluating, for a total of three windows. Because this experiment uses the entire TEQ system at each time step, the best model may change in consecutive months. Furthermore, we apply SHAP~\cite{shap-paper}, which is a feature explainability method, to discover the changes and continuities of important alert content features over time.

\subsection{Triage Module}
\label{meth:sub:triage}

Although the performance over time is informative, it is difficult to understand the impact of the system with machine learning metrics alone. In the triage module, we use the model outputs to present results to analysts in a non-intrusive form. Because intricate interactions with machine learning models and large changes to established workflows are difficult to adopt, we focus our interactions on sorting existing processes. The duality of our model outputs allowing alert-level and incident-level scores grants the capacity to prioritize both incidents as a whole and alerts within a single incident. This results in three strategies, namely:
\begin{itemize}
\item Incident prioritization: ranking of incidents based on their scores to bubble up the most critical incidents to the top of the incident queue.
\item False positive suppression: suppression of incidents with a score that is below an empirically chosen threshold. 
\item Within-incident alert prioritization: ranking alerts within a given incident using the alert-level scores to guide analysts' investigations. 
\end{itemize}

\section{Experiments \& Results}
\label{sec:experimentsresults}

\begin{table*}[t]
\centering % used for centering table
\begin{tabular}{|p{2cm}|p{3.7cm}|*{4}{S|}} 
\cline{3-6}
\multicolumn{2}{c|}{} &\multicolumn{2}{c|}{\textbf{Alerts}} & \multicolumn{2}{c|}{\textbf{Incidents}}\\ 
\specialrule{.13em}{.05em}{.05em} 
Set & Time Span &  \multicolumn{1}{l|}{Positives} &  \multicolumn{1}{l|}{Negatives} &  \multicolumn{1}{l|}{Positives} &  \multicolumn{1}{l|}{Negatives} \\ 
\specialrule{.13em}{.05em}{.05em} 
Training Set 1 & January - June 2022 & 83,156 & 91,843 & 6,188 & 16,629\\
\hline
Test Month 1 & June - July 2022  & 3,522 & 6,489 & 814 & 3,200\\ 
\specialrule{.13em}{.05em}{.05em} 
Training Set 2 & February - July 2022 & 50,488 & 68,684 & 5,678 & 17,026\\ 
\hline
Test Month 2 & July - August 2022 & 14,652 & 11,551 & 753 & 3,075\\ 
\specialrule{.13em}{.05em}{.05em} 
Training Set 3 & March - August 2022 & 45,355 & 65,201 & 4,874 & 17,093\\ 
\hline
Test Month 3 & August - September 2022 & 30,004 & 41,480 & 916 & 2,715\\ 
\specialrule{.13em}{.05em}{.05em} 
\end{tabular}
\caption{Training and test time splits. Every training set is 5 months of data while every test set is one month of held-out data occurring after the training data.}
\label{tab:datasetinfo}
\end{table*}

In this section, we actualize the design decisions explained in the methodology, provide specific details about the experiments, and discuss the results. 

\subsection{Datasets \& Modeling}

The TEQ system is trained and evaluated using real-world data spanning a total of eight months from January 2022 to September 2022, obtained in collaboration with a SOC. There are $282,697$ highly-filtered alerts fired by over $3,185$  sensors on more than $4,054$ different customer estates. These alerts are grouped into $34,279$ incidents. Using the actionable label, $8,668$ incidents are positive and $25,611$ incidents are negative. Because labeling each individual alert is too costly, we propagate the incident-level actionable labels to all alerts within the incident. This results in $131,334$ positive alerts and $151,363$ negative alerts. 

For our modeling experiments, we apply a time split to accurately simulate real-life deployment by training on existing data and evaluating the system on previously unseen future data points. Table~\ref{tab:datasetinfo} summarizes the data splits. The first five months of data are used for training, and the last month is held out for testing. 

In each split, we train content and context models based on various machine learning algorithms (LR, RF, XGB, NN), evaluate them on the corresponding held-out test data on an incident-level, and apply different statistical ensembling strategies to combine the two models into a unified system. We select the model with the highest ROC area under curve (AUC) value. In the following subsections, we detail TEQ's performance over time, then zoom in on the latest test month and present a simulation of TEQ's deployment via our triage results.

\subsection{Performance Over Time}

To evaluate the TEQ framework, we first analyze the performance of the machine learning models over time using traditional machine learning metrics and interpretability methods. 

\subsubsection{Decay Experiments}
\label{subsec:decay}

\begin{figure*}[t]
\centering
   \includegraphics[scale=0.36]{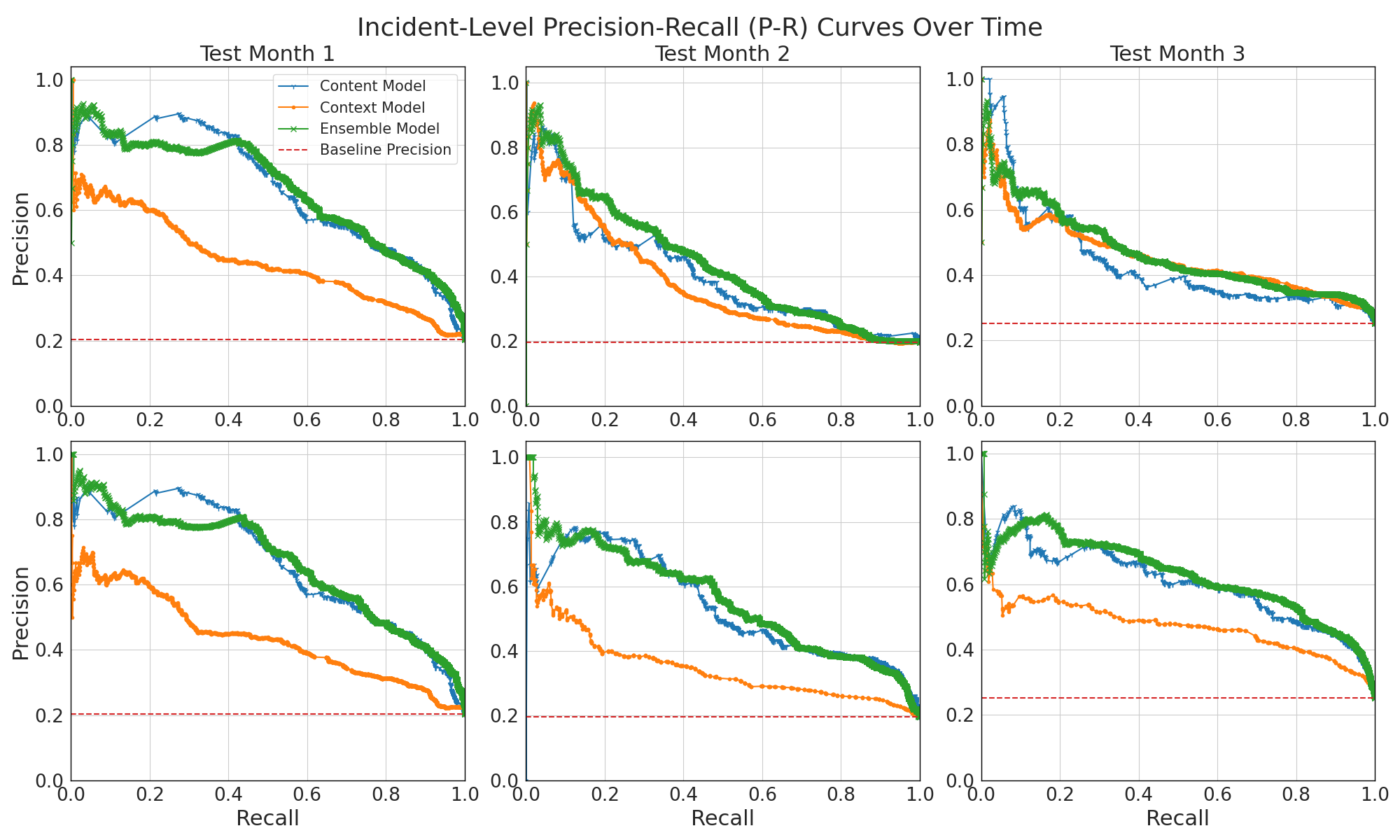}
   \caption{Models' decay over time using precision-recall curves on 5-month training sets. The evaluation is over three consecutive test months under two scenarios: no model retraining (top row) vs. monthly retraining (bottom row). These precision-recall curves show the trade-off between precision and recall for different model thresholds. The recall is the percentage of all actionable incidents that are captured. The P-R AUCs are shown in Table \ref{tab:best-incident-results}.}
   \label{fig:decay_pr} 
\end{figure*}

As detailed in Section~\ref{subsec:temporal-analysis}, we perform two main experiments to study the system's performance decay over time and the impact of retraining. Our findings are summarized in Table~\ref{tab:best-incident-results} and Figure~\ref{fig:decay_pr}. Table~\ref{tab:best-incident-results} shows the best content, context, and ensemble models for every experiment, the selection criteria (ROC AUC) we used for those models, and the specific precision values at a high recall of $95\%$. Notice the lines corresponding to the baseline precision for each test month in Figure \ref{fig:decay_pr}. The baseline precision represents the fraction of actionable incidents in a test set. It shows the SOC's existing state of affairs without TEQ. As shown in Table \ref{tab:best-incident-results}, they correspond to precision values of $20.3\%$, $19.7\%$, and $25.2\%$ for test months 1, 2, and 3, respectively. In contrast, the TEQ ensemble achieves better precision values even at high recall rates.

The top row of Figure~\ref{fig:decay_pr} shows the precision-recall (P-R) curves for the fixed training set decay experiment, i.e. a no retraining setup. In this setup, we simulate the deployment of the best models trained on a fixed 5-month set of data (this is training set 1 in Table~\ref{tab:datasetinfo}) and evaluated one month at a time on the subsequent three months of new, previously unseen data (test months 1, 2, and 3 in Table~\ref{tab:datasetinfo}). The performance on the test set for the first month has an advantage, and there is an obvious decay in the second month. Interestingly, we see a plateau in performance in the third month, rather than a continuation of the decay.The corresponding ROC AUCs for the three test months are $88.1\%$, $70\%$, and $72.9\%$ This could indicate that the model learns both temporally relevant information and long-term trends, and the temporally relevant information is outdated as the threat landscape changes. 

\begin{table*}[!ht]
\centering % used for centering table
\begin{tabular}{|P{1.5cm}|P{0.8cm}|P{4.2cm}|P{2.1cm}*{4}{|P{1.5cm}}|}
\specialrule{.13em}{.05em}{.05em} 
Scenario & Test Month & Task & Model & ROC AUC & Precision-Recall AUC & Precision @ 95\% Recall & Baseline Precision 
\\ 
\specialrule{.13em}{.05em}{.05em} 
\multirow{9}{1.5cm}{\centering Fixed Training Set} & \multirow{3}{0.8cm}{\centering 1} & Content & XGB & 0.871 & 0.669 & 0.332 & \multirow{3}{1.5cm}{\centering 0.203}\\
 \cline{3-7}
 &  & Context & NN & 0.750 & 0.443 & 0.219 & \\
 \cline{3-7}
 &  & Weighted Average  \newline(70\% Content, 30\% Context) & Content: XGB Context: NN & \textbf{0.881} & 0.664 & 0.357 &\\
 \cline{2-8}
  & \multirow{3}{0.8cm}{\centering 2} & Content & XGB & 0.696 & 0.409 & 0.215 & \multirow{3}{1.5cm}{\centering 0.197} \\
 \cline{3-7}
  & & Context & NN & 0.645 & 0.383 & 0.195 &\\
 \cline{3-7}
  & & Weighted Average \newline(70\% Content, 30\% Context) & Content: XGB Context: NN & \textbf{0.700} & 0.444 & 0.200 &\\
 \cline{2-8}
 & \multirow{3}{0.8cm}{\centering 3} & Content & XGB & 0.686 & 0.441 & 0.315 & \multirow{3}{0.8cm}{\centering 0.252}\\
  \cline{3-7}
 &  & Context & NN & 0.728 & 0.456 & 0.305 &\\
 \cline{3-7}
 &  & Weighted Average  \newline(70\% Content, 30\% Context) & Content: XGB Context: NN & \textbf{0.729} & 0.468 & 0.323 &\\ 
\specialrule{.13em}{.05em}{.05em} 
\multirow{9}{1.5cm}{\centering Monthly Retraining} & \multirow{3}{0.8cm}{\centering 1} & Content & XGB & 0.871  & 0.669 & 0.332 & \multirow{3}{0.8cm}{\centering 0.203}\\
  \cline{3-7}
 & & Context & NN & 0.748 & 0.437 & 0.224 &\\
 \cline{3-7}
 & & Weighted Average \newline(70\% Content, 30\% Context) & Content: XGB Context: NN & \textbf{0.881} & 0.666 & 0.356 &\\
 \cline{2-8}
 & \multirow{3}{0.8cm}{\centering 2} & Content & NN & 0.830 & 0.537 & 0.328 & \multirow{3}{0.8cm}{\centering 0.197}\\
 \cline{3-7}
 &  & Context & RF & 0.688 & 0.351 & 0.228 &\\
 \cline{3-7}
 &  &  Weighted Average \newline(70\% Content, 30\% Context) & Content: NN Context: LR & \textbf{0.831} & 0.558 & 0.313 &\\
\cline{2-8}
 & \multirow{3}{0.8cm}{\centering 3} & Content & RF & 0.837 & 0.603 & 0.403 & \multirow{3}{0.8cm}{\centering 0.252} \\
 \cline{3-7}
 & & Context & RF & 0.763 & 0.472 & 0.327 &\\
 \cline{3-7}
 & & Weighted Average \newline(70\% Content, 30\% Context) & Content: RF Context: NN & \textbf{0.852} & 0.627 & 0.411 &\\
\specialrule{.13em}{.05em}{.05em} 
\end{tabular}
\caption{Summary of the incident-level results of the best content and context classifiers, and ensemble for each test month.}
\label{tab:best-incident-results}
\end{table*}

The bottom row of Figure~\ref{fig:decay_pr} demonstrates the outcome of retraining the model throughout the same period using training sets 1, 2, and 3 previously presented in Table~\ref{tab:datasetinfo}. Most notably, the models maintain a strong performance throughout all three timesteps. TEQ's flexibility also shines in this experiment because we can see the best models changing as they are trained on new data as shown in the corresponding scenario in Table~\ref{tab:best-incident-results}. The first month is naturally the same in both experiments with a ROC AUC of $88.1\%$, but in the subsequent two months, retraining the models leads to better performance. Retraining results in a ROC AUC of $83.1\%$ in the second month and $85.2\%$ for the best ensembles. Refer to Tables~\ref{tab:incident-results-tasks} and~\ref{tab:incident-results-ensembles}, and Figure~\ref{fig:decay-rocs} in Appendix~\ref{sec:appendix} for more detailed results for the individual content models, context models, and best performing ensembles. 

\subsubsection{Feature Explainability}

\begin{figure*}[!t]
     \centering
     \begin{subfigure}[b]{0.3\textwidth}
         \centering
         \includegraphics[width=1.0\textwidth]{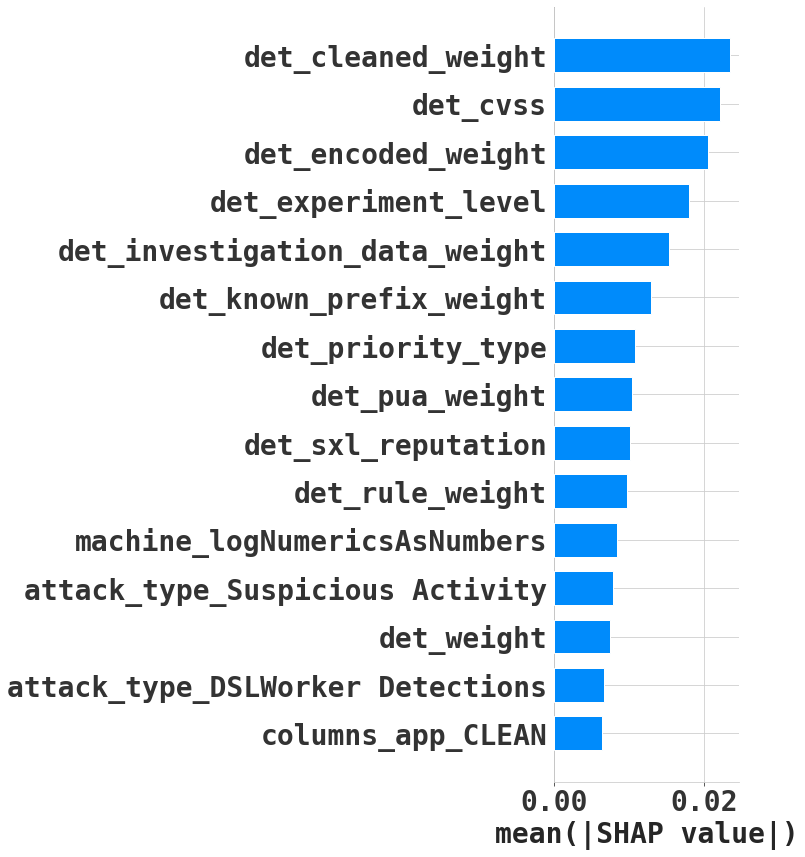}
        \caption{Test Month 1: XGB}
        \label{fig:shap-timestep1} 
     \end{subfigure}
     \qquad
     \begin{subfigure}[b]{0.3\textwidth}
         \centering
         \includegraphics[width=1.0\textwidth]{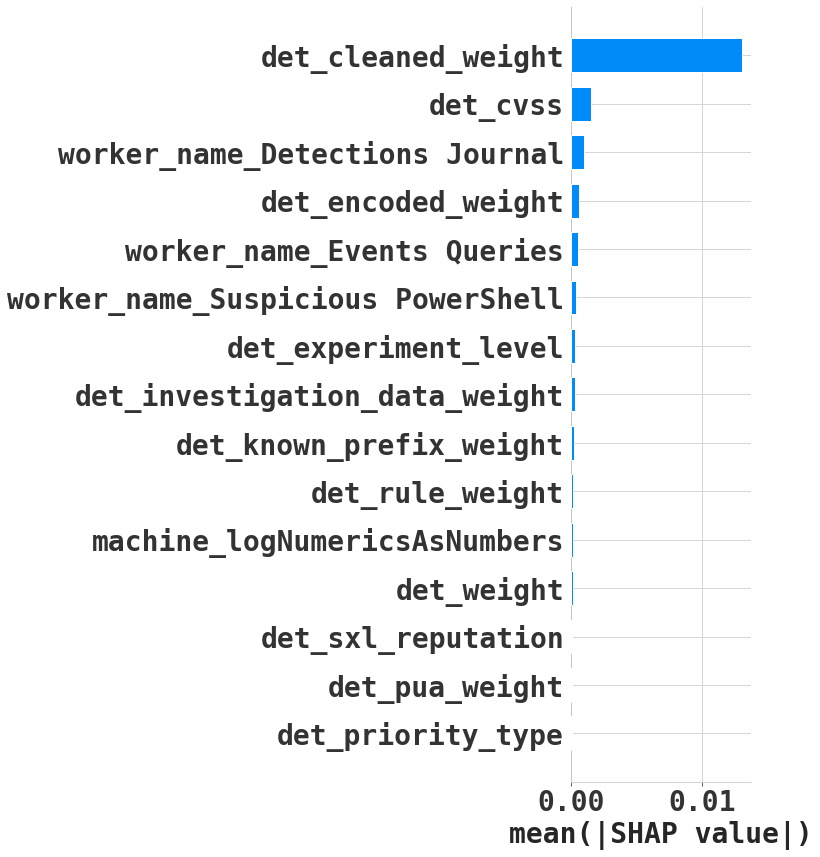}
         \caption{Test Month 2: NN}
         \label{fig:shap-timestep2} 
     \end{subfigure}
     \qquad
     \begin{subfigure}[b]{0.3\textwidth}
         \centering
         \includegraphics[width=1.0\textwidth]{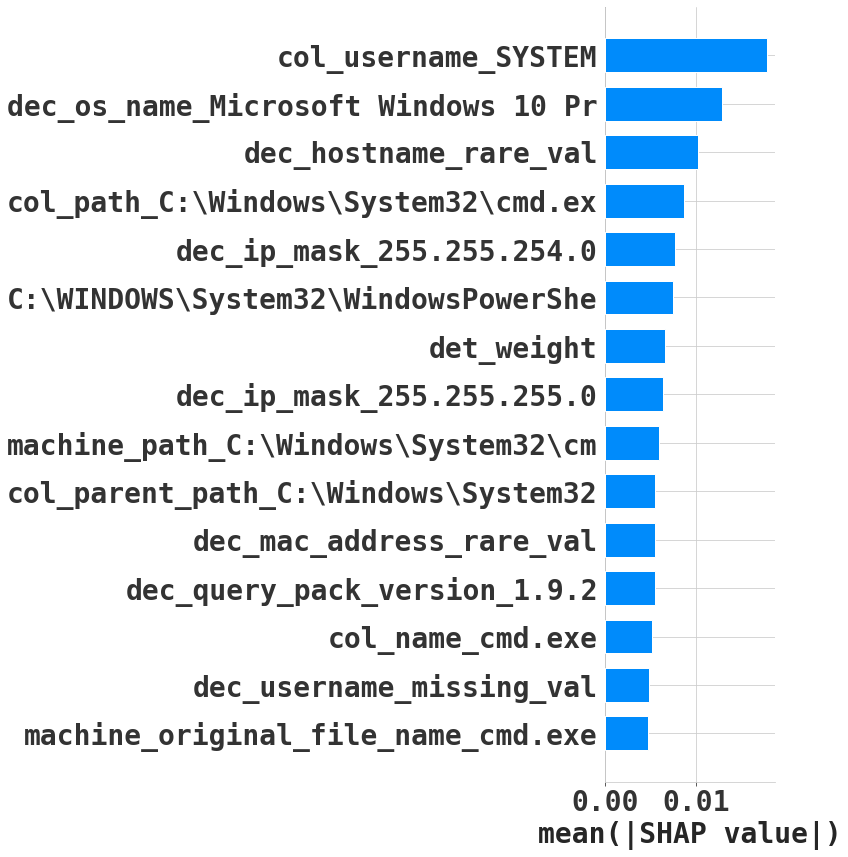}
         \caption{Test Month 3: RF}
         \label{fig:shap-timestep3} 
     \end{subfigure}     
   \caption{SHAP analysis results for the best content models on three test sets in the monthly retraining experiment. For the three test months, the best models are XGB, NN, and RF, respectively. The y-axis contains features from the raw alert contents and the x-axis displays the mean absolute {SHAP} values for those features across all points in the test sets. The mean absolute SHAP value indicates average impact on model output for a given feature. }
   \label{fig:shap-results} 
\end{figure*}

To illustrate the changes and continuities of models over time even further, we perform feature explainability on each test set of our monthly retraining experimental setup. Figure~\ref{fig:shap-results} summarizes the SHAP analysis of the best content models for each of the three timesteps in the retraining experiment displayed in the bottom row of Figure~\ref{fig:decay_pr}. We focus our analysis on the features with the the highest mean absolute SHAP values, which are the average impacts of the features on the models. It is vital to note that these are not exhaustive lists of the features that may influence the models' decisions, and in reality the models are using complex interactions between the features.

In test months 1 and 2, we notice a relatively high overlap in the important features. For test month 1, the classifier is an XGBoost model, and for test month 2, it is a neural network. The models rely heavily on a wide variety of weights, which are all rule-based scoring features to determine the priority of alerts. Despite the high importance of these manual metrics as features, the models still achieve a significant reduction in false positives. This seems to indicate that the breadth of attacks over this time period generally falls within analysts' expectations; however, overall performance can be improved by automatically tuning these manual severity scores, combining features, and doing a deeper inspection of alert content. The models are combining multiple human heuristics to produce better outcomes than the rule-based scores can produce alone.

In test month 3, where the random forest is the best performing classifier, we see a change in the distribution of important features. The difference may be due to a change in the threat landscape. While the rule-based weight is still a relevant feature, a focus on \texttt{cmd.exe}, \texttt{powershell}, and rare host names may indicate a prevalence of attacks on Windows systems in this time period. These features seem to be centered around the characteristics of the events that trigger the alert, which the rule-based scores may be failing to adequately capture. TEQ is able to maintain its performance levels in this time period due to its ability to rapidly retrain models and adapt to information from a massive array of sensors, even if that information is not as useful in previous time periods.

Next, we will dive deeper into the last timestep in the monthly retraining experimental setup as shown in the bottom row Figure~\ref{fig:decay_pr} and showcase TEQ's triage abilities. 

\subsection{Triage Module Results}

To better understand the impact of the TEQ framework, we simulate its deployment over a 1-month held-out test set and measure performance across the three strategies of the triage module described in Section~\ref{meth:sub:triage}: incident prioritization, false positive suppression, and within-incident alert prioritization. For this purpose, we pick the most recent test month in the monthly retraining setup where the model with the highest ROC AUC is the ensemble using a weighted average with a $70\%$ weight for the RF content model and a $30\%$ weight for the NN context model as presented in Table~\ref{tab:best-incident-results}.

\subsubsection{Incident Prioritization}

\begin{figure*}[!htbp]
\centering
   \includegraphics[width=\linewidth]{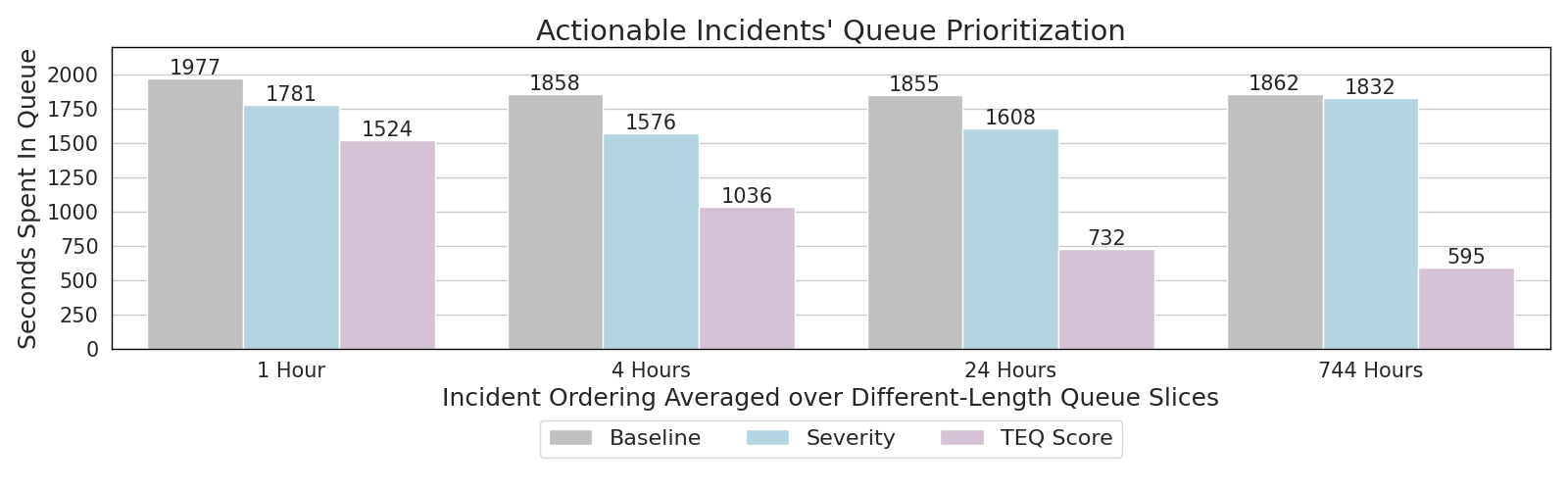}
   \caption{Incident queue prioritization. The bars represent the amount of time actionable incidents sit in queue under different ordering strategies. By reordering the queue according to TEQ incident scores, actionable incidents spend significantly less time in the queue compared to both the analysts' baseline and a severity-based ordering.}
   \label{fig:incident-prioritization} 
\end{figure*}

The first method we employ is the standard incident prioritization that we see across many related work. As incidents are generated, they are chronologically added to a queue for analysts to resolve. In a standard workflow, analysts rely on their domain expertise to pick an incident to investigate first. This means that, inevitably, some incidents will sit in the queue longer before receiving attention. Incident prioritization reorders the queue according to the TEQ incident scores. 

To evaluate the impact of TEQ's incident prioritization, we split our month-long test set based on various time slices, namely: 1 hour, 4 hours, 24 hours, and 744 hours (the full month) in order to demonstrate the effectiveness of TEQ with hypothetical queue/backlog sizes. We then rank the incidents within each slice based on three different criteria: the analyst-selected baseline, the severity scores, which are automatically assigned values that come from hand-written rules, and TEQ incident scores. We sort the real queue times in ascending order and then attach the new incident ordering to those times. This makes sense because we assume that after a specific ordering, analysts take incidents exactly as given according to the ordering. For example, in the 1 hour time slice, we take all the incidents that are created in one hour and order them according to the TEQ incident score. We take the original queue times and sort them in ascending order. The highest ranked incident matches with the smallest queue time. This means that an ordering that places actionable incidents at the top of the queue will reduce the queue times for actionable incidents. 

For each slice and ordering, we average the times that actionable incidents stay in the queue for that slice. Then, we average every slice's mean across the full time span. Figure~\ref{fig:incident-prioritization} summarizes the results of this simulation and shows the impact of TEQ's incident prioritization on the queue times of actionable incidents. Compared to the baseline, we can see that TEQ achieves queue time savings of approximately $22.9\%$, $44.2\%$, $60.5\%$, and $68\%$ for each time slice, respectively. TEQ shows a significant improvement in queue times for actionable incidents in short queues; however, queue sizes vary across SOCs. Although the longer queue sizes are likely not realistic in a real-world SOC, they demonstrate that TEQ's performance increases monotonically, unlike the hand-written rules that generate the severity scores. 

%As a side note, it is possible for actionable incident queue times to decrease because the the non-actionable incident queue times are increasing. We exclude these values from the figure. 

\subsubsection{Incident Suppression}
As an extension to the incident prioritization method, we can apply a classification threshold to TEQ incident scores to completely eliminate incidents from consideration. Because the incident prioritization already ranks all the incidents by the TEQ score, the suppression method is equivalent to dropping incidents from the bottom of the queue. 

The biggest concern that may arise from this strategy is about the few actionable incidents that get removed from the bottom of the queue. At present, most SOCs operate by using hand-written rules and inspecting as many alerts as their human efforts allow. If there are too many alerts to inspect before that effort is expended, then inevitably some actionable incidents will be missed; thus, every SOC has a false negative rate that they must accept. A valuable interpretation of the deployment simulation in the following paragraph is the ratio of actionable to non-actionable incidents in a day. Because the ratio is significantly more advantaged with TEQ in place, a SOC would be able to handle a larger customer base or a greater number of sensors.

\begin{figure*}[!htbp]
\centering
   \includegraphics[width=\textwidth]{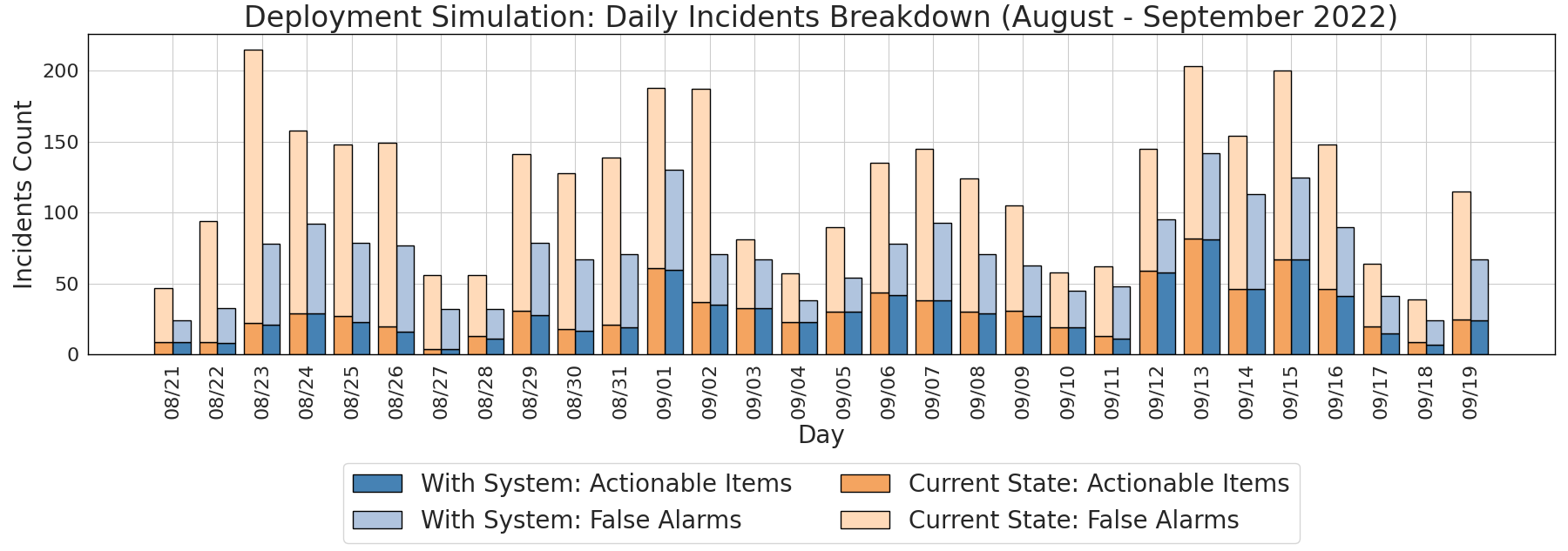}
   \caption{Incident volume over one month of held-out test data. The orange bars represent daily incident resolutions without any information from TEQ. The dark orange sections are the actionable incidents that analysts have to evaluate and the light orange sections are the false positives that the analysts have to evaluate. The blue bars represent the incident resolutions after applying the TEQ system. The dark blue bars represent the amount of actionable incidents that remain and the light blue bars represent the amount of false positives that remain. On average, $54\%$ of false positives are suppressed and $95.1\%$ of actionable incidents are preserved.}
   \label{fig:deployment-plot} 
\end{figure*}

Figure~\ref{fig:deployment-plot} illustrates that daily effect of suppressing incidents using a threshold on top of the the best model in a deployment scenario. Using the P-R curve, we select a classification threshold where the recall is approximately $95\%$. This threshold can be easily tuned, but selecting a high recall value ensures that important incidents are not excluded from analyst evaluation. On the held-out test data, this threshold suppresses $54\%$ of false positives while capturing $95.1\%$ of actionable incidents over the entire month. This deployment scenario shows that it is possible to automatically and consistently reduce false positive incident volume while maintaining a high detection rate.

\subsubsection{Within-Incident Alert Prioritization}

The final triage method we use is within-incident alert prioritization, which helps analysts actually decrease the time it takes to investigate a single incident. This is possible due to the duality of the TEQ framework that allows both incident-level and alert-level scores. In order to evaluate the performance of within-incident alert prioritization, we sampled four incidents and sorted the alerts within each incident in two ways: 1) the standard chronological order, and 2) by TEQ's alert-level score. We provided these to expert analysts for analysis without revealing the origin of the orderings. When relying on TEQ ordering, the analysts were able to determine that a breach occurred in three of the incidents by examining the highest ranked alert and discovering data indicating usage of Meterpreter, Cobaltstrike, and Nishang. With TEQ score ordering, the number of alerts an analyst would have had to investigate is reduced by $14\%$ on average. The chronological order is still important for a holistic understanding of the events that took place on a customer's estate, but the alerts sorted by TEQ's scores can serve as rapid decision making criteria for the actionability of an incident. A dual view between these orderings in a GUI could improve analyst efficiency.

\section{Conclusions}
\label{sec:conclusions}

In this paper, we have demonstrated that TEQ is able to tackle practical issues exacerbating alert fatigue in real-world SOCs. TEQ handles large collections of imprecise sensors, adapts to known false positives even if they are not exact duplicates, incorporates new knowledge from security experts, and offers that knowledge in a non-invasive extension of existing analyst workflows.

The automatic featurization framework shows the feasibility of a hand-off system in large-scale MDR solutions. The featurization system is a key component for integrating numerous organizations and handling chaotic data collection from upstream tasks, allowing those upstreams tasks like sensor modifications and third-party integration to continue their fast-paced nature.

The combination of classic alert contents with contextual features allows the framework to condition on characteristics of different customers - a phenomenon that doesn't occur in small, in-house SOCs because that information doesn't exist. Systems that can adapt to different customer sizes, behaviors, and use cases will be more important as subscription security services grow. 

Our evaluation of TEQ over time uncovers multiple illuminating aspects of the alert prioritization problem that would not have been possible to discover in shorter, singular evaluations, such as TEQ's ability to maintain performance by changing critical features and model types. Due to the rate at which cyberattack strategies are changing, such an evaluation is necessary to gain confidence in the results for realistic data. 

By structuring our models to output alert-level scores with an aggregation scheme for incident-level scores, we can go beyond prioritization and accelerate the investigation process. Completing incidents faster is another way to alleviate alert fatigue that will likely be a prominent topic in future research. 

Finally, our triage module demonstrates $22.9\%$ reduced queue times for actionable incidents, $54\%$ false positive suppression rate, and $14\%$ reduced incident resolution time in a non-invasive workflow. Amidst the increasing rates of cyber crime, complexity of cyberattacks, and the movement to centralized, subscription cybersecurity, a machine learning enabled framework is the key to an effective and scalable solution. 

\section*{Acknowledgment}
Special thanks to Denisz Voznyuk,  Dóra Szabó, and Zoltán Barkó for contributing their domain expertise as cyber analysts. 

\bibliographystyle{unsrt}

\onecolumn
\begin{appendices}
\section{Appendix: Detailed Experimental Results}
\label{sec:appendix}
\setcounter{table}{0}
\renewcommand{\thetable}{A\arabic{table}}
\setcounter{figure}{0}
\renewcommand{\thefigure}{A\arabic{figure}}
\begin{table*}[!ht]
\centering % used for centering table
\begin{tabular}{|P{2.2cm}|P{1.2cm}|P{1cm}|P{1cm}*{6}{|P{1.5cm}}|}
\specialrule{.13em}{.05em}{.05em} 
Scenario & Test Month & Task & Model & ROC AUC  & Precision-Recall AUC & Precision @ 90\% Recall & Precision @ 95\% Recall & Precision @ 99\% Recall & Baseline Precision
\\ 
\specialrule{.13em}{.05em}{.05em} 
\multirow{12}{2.2cm}{\centering Fixed Training Set} & \multirow{8}{1.2cm}{\centering 1} & Content & LR & 0.853 & 0.654 & 0.345 & 0.256 & 0.214 & \multirow{8}{1.5cm}{\centering 0.203}  \\
 \cline{4-9}
 &  &  & RF & 0.845 & 0.615 & 0.328 & 0.295 & 0.225 &\\
 \cline{4-9}
 &  &  &  \textbf{XGB} & \textbf{0.871} & \textbf{0.669} &  \textbf{0.398} &  \textbf{0.332} &  \textbf{0.233} & \\
 \cline{4-9}
 &  &  & NN & 0.862 & 0.658 & 0.368 & 0.336 & 0.232 & \\
 \cline{3-9}
 &  & Context & LR & 0.644 & 0.285 & 0.241 & 0.224 & 0.229 & \\
 \cline{4-9}
 &  &  & RF & 0.707 & 0.423 & 0.217 & 0.211 & 0.208 & \\
 \cline{4-9}
 &  &  & XGB & 0.726 & 0.435 & 0.245 & 0.205 & 0.202 & \\
 \cline{4-9}
 &  &  & \textbf{NN} & \textbf{0.750} & \textbf{0.443} & \textbf{0.268} & \textbf{0.219} & \textbf{0.219} &\\
 \cline{2-10}
  & \multirow{2}{1.2cm}{\centering 2} & Content & XGB & 0.696 & 0.409 & 0.217 & 0.215 & 0.215 & \multirow{2}{1.5cm}{\centering 0.197} \\
 \cline{3-9}
  & & Context & NN & 0.645 & 0.383 & 0.204 & 0.195 & 0.198 & \\
 \cline{2-10}
 & \multirow{2}{1.2cm}{\centering 3} & Content & XGB & 0.686 & 0.441 & 0.323 & 0.315 & 0.269 & \multirow{2}{1.5cm}{\centering 0.252} \\
 \cline{3-9}
 &  & Context & NN & 0.728 & 0.456 & 0.327 & 0.305 & 0.289 & \\
\specialrule{.13em}{.05em}{.05em} 
\multirow{24}{2.2cm}{\centering Monthly Retraining} & \multirow{8}{1.2cm}{\centering 1} & Content & LR & 0.853 & 0.655 & 0.346 & 0.257 & 0.214 &  \multirow{8}{1.5cm}{\centering 0.203} \\
 \cline{4-9}
 & & & RF & 0.846 & 0.614 & 0.316 & 0.292 & 0.223 &\\
 \cline{4-9}
 & & &  \textbf{XGB} &  \textbf{0.871} & \textbf{0.669} &  \textbf{0.398} &  \textbf{0.332} &  \textbf{0.234} &\\
\cline{4-5} \cline{7-10}
 & & & NN & 0.836 & 0.631 & 0.330 & 0.295 & 0.229 &\\
 \cline{3-9}
 & & Context & LR & 0.639 & 0.276 & 0.238 & 0.225 & 0.231 &\\
 \cline{4-9}
 & & & RF & 0.703 & 0.418 & 0.215 & 0.213 & 0.210 & \\
 \cline{4-9}
 & & & XGB & 0.733 & 0.440 & 0.233 & 0.213 & 0.203 &\\
 \cline{4-9}
 & & & \textbf{NN} & \textbf{0.748} & \textbf{0.437} & \textbf{0.277} & \textbf{0.224} & \textbf{0.225} &\\
 \cline{2-10}
 & \multirow{8}{1.2cm}{\centering 2} & Content & LR & 0.790 & 0.498 & 0.299 & 0.256 & 0.219 & \multirow{8}{1.5cm}{\centering 0.197}\\
 \cline{4-9}
 & & & RF & 0.754 & 0.435 & 0.299 & 0.301 & 0.244 &\\
 \cline{4-9}
 & & & XGB & 0.776 & 0.460 & 0.303 & 0.275 & 0.227 &\\
 \cline{4-9}
 & & & \textbf{NN} & \textbf{0.830} & \textbf{0.537} & \textbf{0.365} & \textbf{0.328} & \textbf{0.253} &\\
 \cline{3-9}
 &  & Context & LR & 0.606 & 0.332 & 0.203 & 0.198 & 0.196 &\\
 \cline{4-9}
 & & & \textbf{RF} & \textbf{0.688} & \textbf{0.351} & \textbf{0.251} & \textbf{0.228} & \textbf{0.202} &\\
 \cline{4-9}
 & & & XGB & 0.611 & 0.322 & 0.195 & 0.191 & 0.196 &\\
 \cline{4-9}
 & & & NN & 0.606 & 0.330 & 0.215 & 0.211 & 0.201 &\\
\cline{2-10}
 & \multirow{8}{1.2cm}{\centering 3} & Content & LR & 0.798  & 0.571 & 0.377 & 0.336 & 0.281 & \multirow{8}{1.5cm}{\centering 0.252}\\
 \cline{4-9}
 & & & \textbf{RF} & \textbf{0.837} & \textbf{0.603} & \textbf{0.444} & \textbf{0.403} & \textbf{0.298} &\\
 \cline{4-9}
 & & & XGB & 0.817 & 0.594 & 0.400 & 0.346 & 0.280 &\\
 \cline{4-9}
 & & & NN & 0.826 & 0.591 & 0.404 & 0.355 & 0.283 &\\
 \cline{4-9}
 & & Context & LR & 0.712 & 0.473 & 0.315 & 0.304 & 0.283 &\\
 \cline{4-9}
 & & & \textbf{RF} & \textbf{0.763} & \textbf{0.472} & \textbf{0.360} & \textbf{0.327} & \textbf{0.285} &\\
 \cline{4-9}
 & & & XGB & 0.724 & 0.488 & 0.321 & 0.293 & 0.251 &\\
 \cline{4-9}
 & & & NN & 0.732 & 0.504 & 0.318 & 0.297 & 0.272 &\\
\specialrule{.13em}{.05em}{.05em} 
\end{tabular}
\caption{Summary of the incident-level results of all content and context classifiers for each test month.}
\label{tab:incident-results-tasks}
\end{table*}

\begin{table*}[!htbp]
\centering % used for centering table
\begin{tabular}{|P{1.35cm}|C{0.76cm}|m{4cm}|P{1cm}|P{1cm}|P{0.68cm}|P{0.68cm}*{4}{|P{1.15cm}}|}
\specialrule{.13em}{.05em}{.05em} 
Scenario & Test Month & \centering Ensemble Strategy & Content Model & Context Model & ROC AUC & P-R AUC & Precision @ 90\% Recall & Precision @ 95\% Recall & Precision @ 99\% Recall & Baseline Precision
\\ 
\specialrule{.13em}{.05em}{.05em} 
\multirow{5}{1.4cm}{\centering Fixed Training Set} & \multirow{3}{0.8cm}{\centering 1} & \multirow{2}{4cm}{Weighted Average \newline(70\% Content, 30\% Context)}  & XGB & NN & 0.881 & 0.664 & 0.411 & 0.357 & 0.277 & \multirow{3}{1.15cm}{\centering 0.203} \\
 \cline{4-10}
& & & NN & NN & 0.881 & 0.657 & 0.425 & 0.383 & 0.256 & \\
\cline{3-10}
 & & Average & NN & NN & 0.881 & 0.644 & 0.424 & 0.391 & 0.276 & \\
 \cline{2-11}
  & 2 & Weighted Average \newline(70\% Content, 30\% Context) & XGB & NN & 0.700 & 0.444 & 0.204 & 0.200 & 0.201 & 0.197 \\
 \cline{2-11}
 & 3 & Weighted Average \newline(70\% Content, 30\% Context) & XGB & NN & 0.729 & 0.468 & 0.341 & 0.323 & 0.288 & 0.252\\
\specialrule{.13em}{.05em}{.05em} 
\multirow{10}{1.4cm}{\centering Monthly Retraining} & \multirow{3}{0.8cm}{\centering 1} &  \multirow{3}{4cm}{Weighted Average \newline(70\% Content, 30\% Context)}  & XGB & NN & 0.881 & 0.666 & 0.411 & 0.356 & 0.274 & \multirow{3}{1.15cm}{\centering 0.203}\\
 \cline{4-10}
 & & & XGB & LR & 0.878 & 0.654 & 0.398 & 0.350 & 0.295 &\\
 \cline{4-10}
 & & & XGB & XGB & 0.875 & 0.665 & 0.391 & 0.336 & 0.253 &\\
 \cline{2-11}
 & \multirow{3}{0.8cm}{\centering 2} & \multirow{2}{4cm}{Weighted Average \newline(70\% Content, 30\% Context)} & NN & LR & 0.831 & 0.558 & 0.347 & 0.313 & 0.220 & \multirow{3}{1.15cm}{\centering 0.197}\\
\cline{4-10}
 &  & & NN & RF & 0.829 & 0.553 & 0.334 & 0.291 & 0.219 &\\
\cline{3-10}
 & & Average & NN & LR & 0.823 & 0.560 & 0.332 & 0.266 & 0.216 & \\
 \cline{2-11}
 & \multirow{3}{0.8cm}{\centering 3} & \multirow{2}{4cm}{Weighted Average \newline(70\% Content, 30\% Context)} & RF & NN & 0.852 & 0.627 & 0.450 & 0.411 & 0.297 & \multirow{3}{1.15cm}{\centering 0.252}\\
\cline{4-10}
 & & & RF & RF & 0.851 & 0.627 & 0.439 & 0.403 & 0.315 &\\
 \cline{3-10}
 & & Average & RF & NN & 0.850 & 0.639 & 0.450 & 0.405 & 0.292 &\\ 
\specialrule{.13em}{.05em}{.05em} 
\end{tabular}
\caption{Summary of the incident-level results of the top three ensembling strategies for every test month. In the fixed training set scenario, we only track the best ensemble model in the two subsequent test months.}
\label{tab:incident-results-ensembles}
\end{table*}

\begin{figure*}[!htbp]
\centering
   \includegraphics[scale=0.36]{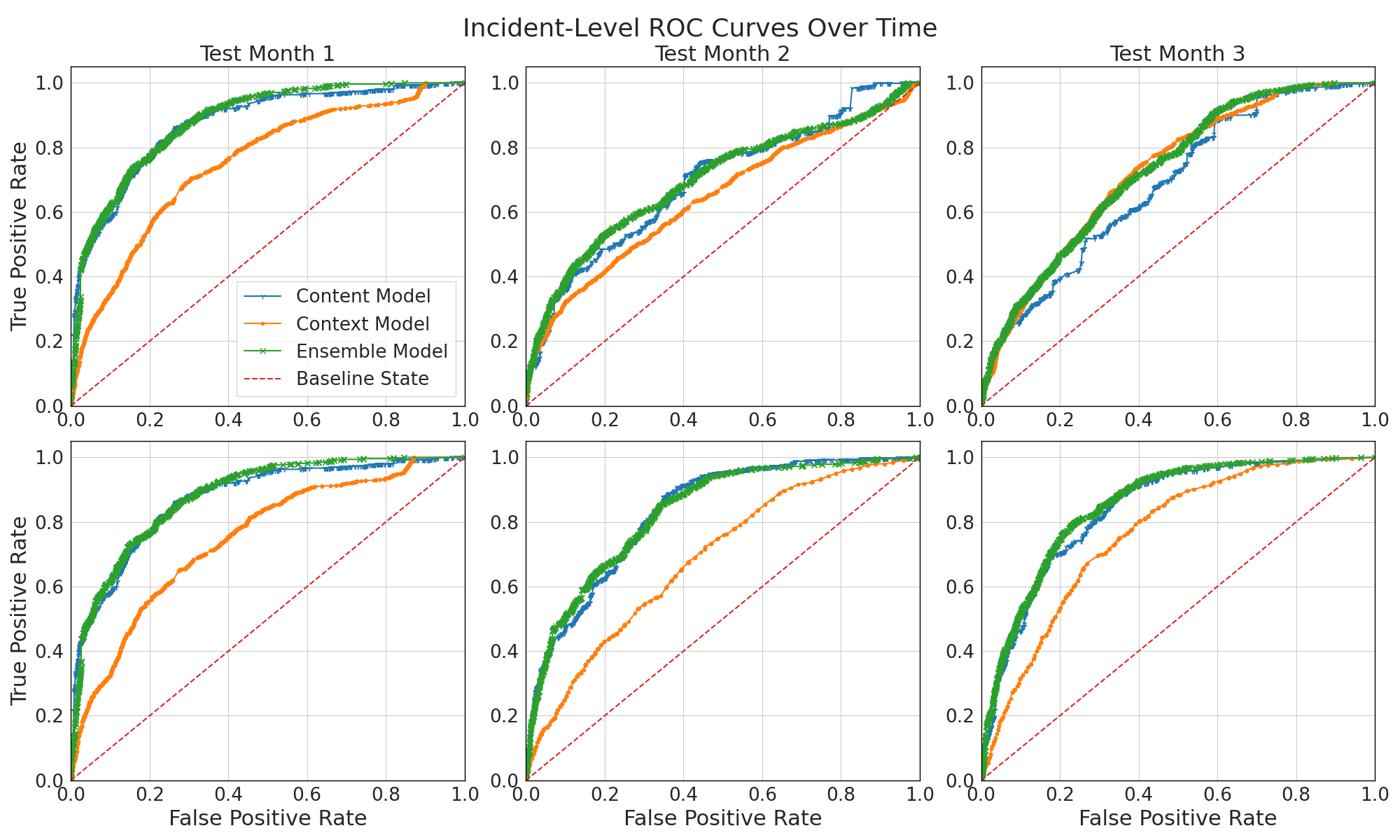}
   \caption{Models' decay over time: Models' ROC curves over three consecutive test months under two scenarios: no model retraining (top row) vs. monthly retraining (bottom row). The corresponding ROC AUCs are shown in Table \ref{tab:best-incident-results}}
   \label{fig:decay-rocs} 
\end{figure*}

\end{appendices}
\end{document}